\documentstyle[12pt]{article}
\def\o{\over}

\def\Bbra{\Big\langle}
\def\Bket{\Big\rangle}

\def\hpt#1{{\tt hep-th/#1}}

\def\ni       {\noindent}
\def\lb       {\left( }
\def\rb       {\right) }
\def\lmb      {\left\{ }
\def\rmb      {\right\} }
\def\lbb     {\left[ }
\def\rbb      {\right] }
\def\comma      { \, , }
\def\period     { \, . }
\def\semiket#1  { \, #1 \, \rangle \, }

\def\half       {  {1\over 2}  }
\def\abs#1      {  \, \vert #1 \vert \,   }
\def\Im#1    { \, {\rm Im } \, #1  }
\def\Re#1    { \, {\rm Re}  \, #1  }
\def\binom#1#2 { \vecii{ {}_{#1} }{\raisebox{.5ex}{$ {}^{#2} $}} }
\def\sqbinom#1#2 { \Bigl(\begin{array}{c} {}_{#1} 
                       \\ \raisebox{.5ex}{${}^{#2}$} \end{array}\Bigr)^2  }

\def\bfR     { {\bf R}}
\def\bfZ     { {\bf Z}}

\def\bfS     { {\bf S}}

\def\calD    { {\cal D} }
\def\calC    { {\cal C} }

\def\calP    { {\cal P} }

\def\zbar   {\bar{z}}
\def\wbar   {\bar{w}}
\def\xbar     {\bar{x}}

\def\mbar     {\bar{m}}
\def\nbar     {\bar{n}}

\def\Jbar   {\bar{J}}
\def\betabar  {\bar{\beta}}
\def\alphabar {\bar{\alpha}}
\def\Pbar {\bar{P}}
\def\Cbar {\bar{C}}
\def\bfS     { {\bf S}}
\def\gammabar  { \bar{\gamma} }
\def\mbar     { \bar{m} }
\def\Up     {\Upsilon}

\def\3F2  { {}_3F_2 }
\def\2F1  { {}_2F_1 }
\def\G#1#2 { G{ #1 \brack #2} }
\def\F#1#2 { F{ #1 \brack #2} }
\def\Fu#1#2#3 { F_{#1}{ #2 \brack #3} }
\def\Gu#1#2#3 { G_{#1}{ #2 \brack #3} }
\def\vecii#1#2      {  \Bigl(\begin{array}{c}#1\\#2\end{array}\Bigr)  }
\def\veciii#1#2#3   {  \left(\begin{array}{c}#1\\#2\\#3\end{array}\right)  }
\def\matrixii#1#2#3#4            {  \Bigl( \begin{array}{cc}#1&#2\\#3&#4
                                       \end{array} \Bigr) }
\def\matrixiii#1#2#3#4#5#6#7#8#9 {  \left(\begin{array}{ccc}#1&#2&#3\\
                                     #4&#5&#6\\#7&#8&#9\end{array}\right)  }
\def\eqb         {  \begin{eqnarray}  }
\def\eqe           {  \end{eqnarray}  }
\def\nn               {  \nonumber  }
\def\sectionnumbering { \setcounter{equation}{0}
         \renewcommand{\theequation}{\arabic{section}.\arabic{equation}}}
\def\appendixnumbering { \setcounter{equation}{0}
         \renewcommand{\theequation}{\Alph{section}.\arabic{equation}}}
\def\mysection#1{ \addtocounter{section}{1} \setcounter{subsection}{0}
                 \sectionnumbering 
   \par \bigskip
      \par \bigskip \noindent
   {\bf \arabic{section} \quad  #1 } 
    \par \bigskip}
\def\appsection#1{\addtocounter{section}{1} \setcounter{subsection}{0}
                 \appendixnumbering
    \par \bigskip 
      \par \bigskip \noindent
   {\bf \Alph{section} \quad  #1 } 
    \par \bigskip}   
\def\mysubsection#1{\addtocounter{subsection}{1}
      \par \bigskip \noindent  {\normalsize\it
      \arabic{section}.\arabic{subsection} \quad #1  }
   \par \medskip }

\def\csectionast#1    { \begin{center}
    {\large\bf #1  }   \end{center} \par \bigskip}
\def\titleandfile#1#2   {  \begin{center}{\large\bf #1}\end{center}
                            \par\begin{flushright} #2 \end{flushright}  }
\renewcommand{\thefootnote}{\fnsymbol{footnote}}

\begin{document}
%%%%%%%%%%%%%%%%%%%%%
%    cover
%%%%%%%%%%%%%%%%%%%%%
\thispagestyle{empty}
\setcounter{page}{0}
%%%%% pre-print number %%%%%

\baselineskip 5mm
\renewcommand{\thefootnote}{\fnsymbol{footnote}}
\hfill\vbox{
\hbox{UTHEP-446}
\hbox{hep-th/0109059}
}

%%%%% title %%%%%

\baselineskip 0.8cm
\vskip 30mm
\begin{center}
{\large\bf Three-point functions and operator product expansion \\
in the $ SL(2)$ conformal field theory}
\end{center}

%%%%% authors %%%%%

\vskip 17mm
\baselineskip 0.8cm
\begin{center}
    Yuji ~Satoh\footnote[2]
    {\tt ysatoh@het.ph.tsukuba.ac.jp}                           \\
       {\it Institute of Physics, University of Tsukuba}        \\
       {\it Tsukuba, Ibaraki 305-8571, Japan}
\end{center}

%%%%% abstract %%%%%

\vskip 30mm
\baselineskip=4ex
\begin{center}{\bf Abstract}\end{center}
\par
\bigskip

In the $SL(2)$ conformal field theory,   
we write down and analyze the analytic expression of the three-point 
functions of generic primary fields with definite $SL(2)$ weights.
Using these results, we  
discuss the operator product expansion in the $SL(2,R)$ WZW model.
We propose a prescription of the OPE, the classical limit of which 
is in precise agreement with the tensor products 
of the representations of $SL(2,R)$.

%
%\vskip 5mm

\noindent 
%{\sl PACS}:~ 11.25.-w, 11.25.Hf \\
\noindent
%{\sl keywords}:
%
\vfill
\noindent
September~ 2001

\newpage
%%%%%%%%%%%%%%%%%%%%%%%%
% Main text
%%%%%%%%%%%%%%%%%%%%%%%%

\renewcommand{\thefootnote}{\arabic{footnote}}
\setcounter{footnote}{0}
\setcounter{section}{0}
\baselineskip = 0.6cm
\pagestyle{plain}

%%%%%%%%%%%%%%%%%%%%%%%%
\mysection{Introduction}
%%%%%%%%%%%%%%%%%%%%%%%%
%
The conformal field theory with the $\widehat{sl}(2)$ symmetry 
provides us with some of the simplest models beyond the well-studied
rational CFT. It is an interesting subject by itself. Moreover, 
since the $SL(2)$ symmetry is rather general, 
it appears in various situations, for instance, 
in studying the strings on $AdS_3$, black holes in string theory and
certain problems in condensed matter physics. 
However, there still remain open questions 
about the models with the $\widehat{sl}(2)$ symmetry. 

In this paper, we would first like to discuss the three-point functions
in the $SL(2)$ conformal field theory. 
One of the models with the $ \widehat{sl}(2)$ symmetry is 
the $H_3^+ $ WZW model \cite{Gawedzki,Teschner1,Teschner2}.
The primary fields which are often used in this model 
are labeled by the $SL(2)$ spin $j$ as $\Phi_j$.
For these primary fields, the two- and three-point functions 
including normalizations have been obtained 
\cite{Teschner1,Teschner2} (see, also \cite{Ishibashi-OS,Hosomichi-OS}).
Since those expressions are analytic
in $j$'s (up to delta functions), 
one expects that the correlation functions in other $SL(2)$
models may be obtained by appropriately continuing the values of the spin.
However, in a model where highest(lowest) weight representations
appear, it is important to respect 
the relations between the spin and the $SL(2)$ weights. Thus, one needs 
the primary fields with definite left and right $SL(2)$ weights, 
 $\Phi^j_{m\mbar}$. An emphasis has been put on this point 
in \cite{Hosomichi-S}.
With the help of an integral formula in \cite{Fukuda-H}, in section 2 
we write down the analytic expression of the three-point 
functions of $\Phi^j_{m\mbar}$ in the $H_3^+$ WZW model. 
Some properties of these three-point functions are also analyzed.

Based on the results in section 2, we would next like to discuss the
operator product expansion in the $SL(2,R)$ WZW model. 
For this model, it has been a long-standing problem to 
determine the correct spectrum (see, e.g., 
\cite{BOFW}-\cite{Hikida-HS} and references therein). 
The OPE gives us an important clue. 
Guided by a consideration in the classical limit, in section 3 we propose 
the OPE of the primary fields in the $SL(2,R)$ WZW model. In the classical 
limit, we find 
a complete agreement with the classical tensor products of 
the representations of $SL(2,R)$. Rather, the full OPE
is essentially the same as the classical tensor products. This is 
natural because one may not have any reason 
that makes difference, contrary to the $SU(2)$ case. 
Although we take a different strategy to determine the OPE, the 
resultant prescription is regarded as a completion of the arguments in 
\cite{Hosomichi-S}. 

Implications to the spectrum of the $SL(2,R)$ WZW model are briefly 
discussed in section 4, in particular, 
in relation to the works in \cite{Maldacena-O}.  
Some formulas used in the main text and some technical arguments 
are given in the appendix.

%
%\vspace*{-2ex}
\newpage
%%%%%%%%%%%%%%%%%%%%%%%%%
\mysection{Three-point functions}
%%%%%%%%%%%%%%%%%%%%%%%%
%
\vspace*{-3ex}
\mysubsection{$H_3^+$ WZW model}
We begin with a brief summary of the correlation functions 
in the  $H_3^+$($=SL(2,C)/SU(2)$) WZW model. We mainly follow the 
notations in \cite{Hosomichi-S}.\footnote{ 
However, we use slightly different notations. 
For example, $\Phi^j_{m\mbar}$,  
$(2\pi)^2 \delta^2(\Sigma m_a)$ and $W(j_a;m_a)$ correspond 
to $\Phi^{-j-1}_{m\mbar}$, $ \pi \delta^2(\Sigma m_a)$ and 
$F(-j_a-1;m_a)$ in \cite{Hosomichi-S}, respectively.
}

The conformal field theory with the Euclidean $AdS_3$ target space is 
described by the $H_3^+$ WZW model \cite{Gawedzki,Teschner1,Teschner2}. 
This model has an $SL(2,C) \times \overline{SL}(2,C)$ affine symmetry, 
whose currents act on the primary fields as
\eqb
 J^a(z)\Phi_j(w,x)
     &\sim&  -\frac{D^a\Phi_j(w,x)}{z-w} \comma \label{sl2Phi}
\eqe
$$
  D^- \ =  \   \partial_x  \comma \quad 
  D^3 \ = \ x  \partial_x -  j  \comma \quad 
  D^+ \ = \ x^2\partial_x - 2jx \comma
$$
and similarly for $\Jbar^a(\zbar)$. $z,w$ are the world-sheet
coordinates. $x$ is a complex parameter. The two- and three-point 
functions of $\Phi_j$ have been calculated by using symmetries 
\cite{Teschner1,Teschner2}, or later by path-integral 
\cite{Ishibashi-OS,Hosomichi-OS}:
\eqb
  && \Bbra \Phi_{j_1}(z_1,x_1)\Phi_{j_2}(z_2,x_2) \Bket \label{2ptPhi} \\
  && \quad 
  \ = \
 |z_{12}|^{-4h_1}\Bigl[ A(j_1)\delta^2(x_{12}) \delta(j_1+j_2+1)
                          +B(j_1)|x_{12}|^{4j_1} \delta(j_1-j_2) 
  \Bigr]
 \comma \nn \\
 && \quad \qquad A(j) \ = \ -\frac{\pi^3}{(2j+1)^2} \comma \quad 
 B(j) \ = \ b^2\pi^2[k^{-1}\Delta(b^2)]^{2j+1}\Delta[-b^2(2j+1)] \comma \nn \\
  &&  \Bbra \prod_{a=1}^3\Phi_{j_a}(z_a,x_a) \Bket
 \ = \  D(j_a)\prod_{a<b}|z_{ab}|^{-2h_{ab}}|x_{ab}|^{2j_{ab}} \comma 
   \label{3ptPhi} \\
 && \quad \qquad D(j_a) \ = \
  \frac{b^2\pi}{2}
  \frac{[k^{-1}b^{-2b^2}\Delta(b^2)]^{\Sigma j_a+1}
        \Upsilon[b]\Upsilon[-2j_1b]\Upsilon[-2j_2b]\Upsilon[-2j_3b]}
       {\Upsilon[-(\Sigma j_a+1)b]\Upsilon[-j_{12}b]\Upsilon[-j_{13}b]
        \Upsilon[-j_{23}b]}
 \period \nn
\eqe
Here, 
$x_{ab} = x_a - x_b$, $z_{ab} = z_a - z_b$, 
$j_{12} = j_1 + j_2 - j_3$, $h_{12} = h_1 + h_2 -h_3$ etc., 
$h_a = -j_a(j_a+1)b^{2}$, 
$b^{-2} = k-2$, and $\Delta(x) = \Gamma(x)/\Gamma(1-x)$. 
$k$ is the level of $\widehat{sl}(2)$. $\Up(x)$ 
is a certain function introduced in \cite{Zamolodchikov-Z}.  

In this paper, we are interested in the primary fields with 
definite $SL(2)$ weights. They are given by the moments with respect 
to $x$:
\eqb
     \Phi^j_{m\mbar}(z) & = & \int d^2x \,
  x^{j+m}\xbar^{j+\mbar} \Phi_{-j-1}(z,x)
   \period  \label{Phijmm}
\eqe
The currents act on $\Phi^j_{m\mbar}$ as
\eqb
  J^\pm(z) \Phi^j_{m\mbar}(w) &\sim&  \frac{\mp j +m}{z-w} 
     \Phi^j_{m\pm 1 \, \mbar}
    \comma  \quad 
    J^3(z) \Phi^j_{m\mbar}(w) \ \sim \  \frac{m}{z-w} \Phi^j_{m\mbar}
  \comma \label{JPhijmm}
\eqe
and similarly for $\Jbar^a(\zbar)$.
Here and in the following, we assume that $ m - \mbar \in \bf Z$, 
so that the above integral is well-defined. This is 
the case for the unitary representations on $H_3^+$ (and $SL(2,R)$).  

From (\ref{2ptPhi})-(\ref{Phijmm}),  
the correlation functions of $\Phi^j_{m\mbar}$ are expressed as 
\cite{Giveon-K,Maldacena-O,Giribet-N2,Hosomichi-S},
\eqb
   && \Bbra \Phi^{j_1}_{m_1\mbar_1} \Phi^{j_2}_{m_2\mbar_2}\Bket
        \label{Phijm2pt} \\ 
    && \quad    
     = \ (2\pi)^2 \delta^2(m_1+m_2) 
        \lbb A(j_1) \delta(j_1+j_2+1) 
    + B_\Phi(j_1,m_1) \delta(j_1-j_2) \rbb \comma \nn  \\
   && \qquad 
   B_\Phi(j,m) = c^{-j-1}_{m\mbar} B(-j-1) \comma \quad 
      c^{j}_{m\mbar} = \pi 
     {\Delta(2j+1) \Gamma(-j+m)\Gamma(-j-\mbar) 
     \o \Gamma(j+1+m) \Gamma(j+1-\mbar)}
     \comma \nn \\
   && \Bbra \prod_{a=1}^3 \Phi^{j_a}_{m_a\mbar_a} \Bket
   = (2\pi)^2 \delta^2(\Sigma m_a) W(j_a;m_a) D(-j_a-1)
  \comma \label{Phijm3pt}
\eqe
where
\eqb
  W(j_a;m_a) & = & \int d^2x_1 d^2x_2 \ 
     x_1^{j_1+m_1}\xbar_1^{j_1+\mbar_1} |1-x_1|^{-2j_{13}-2}  
      \label{intWjm}  \\
     & & \qquad \times \  
     x_2^{j_2+m_2} \xbar_2^{j_2+\mbar_2} |1-x_2|^{-2j_{23}-2}
     |x_1-x_2|^{-2j_{12}-2}
    \period  \nn
\eqe
In the above, we have omitted the $z$-dependence. 
Note that $c^j_{m\mbar} = c^{j}_{\mbar m}$ for $ m-\mbar \in \bfZ $.
\mysubsection{Three-point functions of $\Phi^j_{m\mbar}$}
In a special case with $j_1+m_1=j_1+\mbar_1 = 0$, the integral
in (\ref{intWjm}) has been carried out in \cite{Hosomichi-S}.
For the case of $m_a = \mbar_a$, see also 
\cite{Becker-B,Giribet-N2}. In this subsection, we would like to 
write down the explicit form of $W(j_a;m_a)$ in a generic case.

The calculation proceeds as follows. One can `factorize' the integration
over the complex variables, $x_1,x_2$, as in \cite{Dotsenko-F}. Then, 
an integration over one variable gives $ \2F1 $, and the remaining
integration gives $ \3F2 $, where ${}_pF_q$ are the (generalized)
hypergeometric functions. The final result is expressed by certain
combinations of $ \3F2 $. In this course, one needs careful 
treatment of the integration contours. 

Such calculations have been performed in \cite{Fukuda-H}. Thus, we have
only to use the simplest formula there given in (\ref{I1}), to find that 
\eqb
  W(j_a;m_a) &=& (i/2)^2 \lbb C^{12} \Pbar^{12} + C^{21} \Pbar^{21} \rbb
  \comma \label{Wjm1}
\eqe
where
\eqb
  (i/2)^2 P^{12} &=& s(j_1+m_1) s(j_2+m_2) C^{31} 
         - s(j_2+m_2) s(m_1-j_2+j_3) C^{13} \comma  \nn  \\
  C^{12}
     &=& {\Gamma(-N) \Gamma(1+j_3-m_3) \o \Gamma(-j_3-m_3)}
         \G{-j_3-m_3, -j_{13}, 1+j_2+m_2}{-j_1+j_2-m_3+1, -j_1-j_3+m_2}
     \comma \label{PCC}  \\
   C^{31} &=& { \Gamma(1+ j_3 + m_3) \Gamma(1 + j_3 - m_3) \o
                       \Gamma(1+N)}
           \G{1+N,1+j_1+m_1,1+j_2-m_2}{j_2+j_3+m_1+2,j_1+j_3-m_2+2}
   \comma \nn
\eqe
$s(x) =  \sin (\pi x)$, and 
$N = j_1+j_2+j_3+1$. $G$ is defined by 
$ \G{a,b,c}{e,f} = {\Gamma(a) \Gamma(b) \Gamma(c) \o \Gamma(e) \Gamma(f)}
    \F{a,b,c}{e,f} $ with $ \F{a,b,c}{e,f} = \3F2 (a,b,c;e,f;1) $.
The bar in $\Pbar^{12} (\Pbar^{21})$ means to replace $m_a$ in 
$P^{12} (P^{21})$ with $\mbar_a$, i.e., 
$\Pbar^{ab} \ = \ P^{ab} (m_a \to \mbar_a)$. 
$C^{21}, C^{13}, P^{21}$ are obtained   
by exchanging $j_1,m_1$ and $j_2,m_2$ in $C^{12}, C^{31}, P^{12}$,
respectively: 
\eqb
 C^{ab}  &= &  C^{ba} (j_1,m_1 \leftrightarrow j_2,m_2) \comma 
   \quad 
 P^{ab}  \  = \ P^{ba} (j_1,m_1 \leftrightarrow j_2,m_2)
 \period
\eqe  
We have also used $ \Sigma_a m_a = \Sigma_a \mbar_a = 0$. 
The expression of $W(j_a;m_a)$ is analytic in its arguments, and 
so are the three-point functions of $\Phi^j_{m\mbar}$ except for 
$\delta^2(\Sigma m_a)$.

For later use, we further rewrite $W(j_a;m_a)$. 
As discussed in  appendix B, it turns out that $P^{12}$ and $P^{21}$
are expressed in terms of  $C^{12}$ and $C^{21}$.
From the formula (\ref{I2}), it follows that
\eqb
   W(j_a;m_a) &=& D_1 C^{12} \Cbar^{12} + D_2 C^{21} \Cbar^{21}
             + D_3  \lbb C^{12} \Cbar^{21} +C^{21} \Cbar^{12} \rbb
  \comma \label{Wjm2}
\eqe 
where
$\Cbar^{ab} = C^{ab}(m_a \to \mbar_a)$ and 
\eqb
   D_{1} &=& {s(j_2+m_2) s(j_{13}) \o s(j_1-m_1) s(j_2-m_2) s(j_3+m_3)}
        \lbb \, s(j_1+m_1) s(j_1-m_1) s(j_2+m_2) \right. \nn \\
   && \  \qquad \left. 
       -  \ s(j_2-m_2) s(j_2-j_3-m_1) s(j_2+j_3 -m_1) \, \rbb \comma \nn \\
   D_{2} &=& D_{1} (j_1,m_1 \leftrightarrow j_2,m_2) \comma  \label{DC} \\
   D_3 &=& -{s(j_{13}) s(j_{23}) s(j_1+m_1) s(j_2+m_2) s(j_1+j_2+m_3) 
            \o s(j_1-m_1) s(j_2-m_2) s(j_3+m_3)}  
   \period \nn 
\eqe
Since $m_a - \mbar_a \in \bfZ$, (\ref{Wjm2}) is symmetric with respect to
$m_a$ and $\mbar_a$, as it should be. 

So far, we have discussed the correlation functions 
in the $H_3^+$ WZW model, where the parameters in 
$\Phi^j_{m\mbar}$ have been  $j = -1/2 + i \rho$, $m = \half(ip+q)$, 
$\mbar = \half(ip-q)$ with $\rho, p \in \bfR $, $q \in \bfZ$.
In the following, we would like to assumed that the correlation functions
in other models with an $ \widehat{sl}(2) \times \widehat{sl}(2) $ symmetry 
are obtained by appropriately continuing the parameters in the above results 
\cite{Teschner1,Teschner2,Giveon-K,Maldacena-O,Giribet-N2,Hosomichi-S}.
In particular, we assume that the correlation functions in the $SL(2,R)$ WZW 
model are obtained by setting $j,m,\mbar$ to be the values of the 
representations of $SL(2,R)$. 

To understand this, we first note that, from the algebraic point of view, 
the symmetries of the $H_3^+$ and $SL(2,R)$ WZW models are the same, i.e., 
$ \widehat{sl}(2) \times \widehat{sl}(2) $. The difference is the allowed 
representations and, hence, the allowed values of the parameters, $j, m$ and 
$\bar{m}$. In addition, the correlation functions discussed so far respect 
the $ \widehat{sl}(2) \times \widehat{sl}(2) $ symmetry, and they are 
analytic in $j, m $ and $\bar{m}$ (up to delta functions). 
Then, the assumption follows from these facts.

In fact, we will find that, after taking into account the difference of 
the definitions of the operators, the continued correlation functions 
correctly reduce to the known ones for the $SL(2,R)$ WZW model 
\cite{Becker-B} (up to phases) in a special case of (\ref{W10}) with 
$m_a = \mbar_a$.
Furthermore, we can find more evidence for the validity of such 
a continuation: (i) As discussed in \cite{Teschner1} and \cite{Hosomichi-S},
by using the continued correlators, one can obtain the correct OPE 
of the models with the $ \widehat{sl}(2) \times \widehat{sl}(2) $ symmetry 
other than the $H_3^+$ WZW model. 
(ii) As we will discussed in section 3, the continued correlation functions 
have the correct pole structure which is required from the representation 
theory of $SL(2,R)$. This is quite non-trivial, and implies that they are 
actually regarded as the correlators of the $SL(2,R)$ WZW model. 
It also turns out that the OPE based on these correlators is consistent with 
the $SL(2,R)$ symmetry. 
(iii) If we adopt, e.g., the path integral approach, it is not possible 
to directly obtain the correlators of the $SL(2,R)$ WZW model; 
the integral diverges because of the time like direction. Thus, 
in analogy to the case of the flat target space-time, it is natural 
to `define' the correlators of the $SL(2,R)$ case by some continuation 
from those in the Euclidean case, i.e., the $H_3^+$ case. 
Such an issue has also been discussed in detail in a recent paper 
\cite{Maldacena-O2}. 
(iv) Some correlators for other models which are obtained 
by such a continuation from those in the $H_3^+$ case have been 
used in the literature, e.g., \cite{Maldacena-O,Giveon-K},  
and the results seem to be physically reasonable.

As is discussed shortly, our results 
pass some other consistency checks in addition to those mentioned above.  
Therefore, our assumption has fairly good grounds. 

In the following, among the $SL(2,R)$ representations, we concentrate 
on the normalizable unitary representations, whose parameters are
given by
\par\medskip\ni
\quad 
(1) \  
  $j < -1/2$, $m = -j + \bfZ_{\geq 0}$ \ for the lowest weight
    discrete series ($\calD^+_j$);   
\par\ni
\quad 
(2) \ 
  $j < -1/2$, $m = j - \bfZ_{\geq 0} $  \ for the highest weight
    discrete series ($\calD^-_j$);
\par\ni
 \quad 
(3) \  
   $j = -1/2 + i \bfR_{> 0}$, $ m = \alpha + \bfZ $
   ($ 0 \leq \alpha < 1 $) \ for the principal continuous series 
    ($\calC^\alpha_j$).
\par\medskip\ni
$\mbar$ takes the values in the same representation as $m$, because 
we consider the case where $\Phi^j_{m\mbar}$ give the diagonal 
combinations of the left and right $SL(2,R)$ representations.
In this case, $\Phi^j_{m\mbar}$ correspond to 
the matrix elements of the above three types of representations, 
which span the $L^2$-space on an $SL(2,R)$ manifold. 

%\newpage
%
\mysubsection{Case with highest(lowest) weight representations}
When $j$ and $m$ satisfy a certain relation, the $\widehat{sl}(2)$ 
representation largely reduces because of the appearance 
of the null vectors. The correlation functions 
in such a case may also be quite different from those in a generic case.
Given (\ref{Wjm1}) and (\ref{Wjm2}), we can confirm this.

As an example, we consider the case in which $\Phi^{j_1}_{m_1\mbar_1}$
belongs to a left and right combination of lowest weight representations: 
\eqb
   &&  m_1 = -j_1 + n_1 \comma \quad \mbar_1 = -j_1 + \nbar_1 
   \qquad (n_1, \nbar_1 \in \bfZ_{\geq 0}) \period \label{lowest} 
\eqe    
$D_2$ and $D_3$ in (\ref{DC}) then vanish
when the parameters are generic except for satisfying (\ref{lowest}).
In addition, the analysis
in appendix C shows that $C^{12}$ and $C^{21}$ do not give 
divergences which cancel the zeros in $D_{2}$ and $D_3$.
Thus, only the first term in (\ref{Wjm2}) remains non-vanishing, 
so that  $W(j_a;m_a)$ reduces to 
\eqb
  &&  W_1 (j_a;m_a) =  (-)^{m_3-\mbar_3+\nbar_1} 
    {\pi^2 \Delta(-N) \Delta(2j_1+1) \o \Delta(1+j_{12}) \Delta(1+j_{13})}
 {\Gamma(1+j_3 -m_3) \Gamma(1+j_3-\mbar_3) \o 
  \Gamma(1+j_3-m_3-n_1) \Gamma(1+j_3-\mbar_3-\nbar_1)} 
  \nn    \\
  &&   \qquad \times \   
   \prod_{a=2,3} {\Gamma(1+j_a+m_a) \o \Gamma(-j_a-\mbar_a)}  
    \F{-n_1,-j_{12},1+j_{23}}{-2j_1,1+j_3-m_3-n_1}
    \F{-\nbar_1,-j_{12},1+j_{23}}{-2j_1,1+j_3-\mbar_3-\nbar_1}
  \period \nn \\
  &&  \label{W1}
\eqe 
Here, we have used the first expression of $C^{12}$ in (\ref{Cs}).
Note that $F$'s above are finite sums, since they have non-positive
integers in their upper arguments.

This is in accord with the discussion
in \cite{Holman-B}: $W(j_a;m_a)$ is essentially 
the left and right combination of the $SL(2)$ Clebsch-Gordan coefficients.
Generically, there are two linearly independent solutions to 
the difference equation for each Clebsch-Gordan coefficients. 
However, when
one of the operator is set to be in a highest(lowest) weight 
representation, only one solution remains because of boundary conditions.
$C^{12}$ is regarded as such a solution and, hence, only 
the combination $D_1 C^{12} \Cbar^{12}$ remains.

If we further set $n_1 = \nbar_1 = 0$ in (\ref{W1}), 
the $F$'s above become unity. Consequently, we obtain
\eqb
  W_1^0(j_a;m_a) & = & (-)^{m_3-\mbar_3} \pi^2 
    {\Delta(-N) \Delta(2j_1+1) \o \Delta(1+j_{12}) \Delta(1+j_{13})}
     \prod_{a=2,3} {\Gamma(1+j_a+m_a) \o \Gamma(-j_a-\mbar_a)}
   \period \label{W10}
\eqe 
This is in precise agreement with the result in \cite{Hosomichi-S}, 
including the phase.

Since $\Phi^{j}_{m\mbar}$ form $SL(2)$ representations with respect to
the zero-modes of the currents, $J^a_0$ and $\Jbar^a_0$, it should 
be possible to derive (\ref{W1}) conversely from (\ref{W10}). 
One can show this following \cite{Becker-B}. 
In appendix D, we sketch the corresponding argument in our case.
We remark that (\ref{W1}) does not 
coincide with the corresponding quantity in \cite{Becker-B} even 
when $m_a = \mbar_a$.
This is because the phase of 
$W_1^0(j_a;m_a)$ is different, and gives a different
analytic expression in a generic case.

%
%%%%%%%%%%%%%%%%%%%%%%%%
\mysection{OPE in the $SL(2,R)$ WZW model}
%%%%%%%%%%%%%%%%%%%%%%%%
%
In this section, we would like to discuss the OPE of the primary 
fields in the $SL(2,R)$ WZW model. As mentioned in 
section 2, we suppose that 
the correlations functions in the $SL(2,R)$ WZW model are obtained
from those in section 2. 
It is also understood that the parameters take generic values
unless otherwise stated. The cases with special values are 
obtained by taking appropriate limits. In addition, when singular 
quantities appear, we regularize them by infinitesimally 
shifting the parameters, and take limits in the final expressions.

\mysubsection{Preliminary}

We start with the following general form of the OPE among 
$\Phi^{j}_{m\mbar}$:
\eqb
 \Phi^{j_1}_{m_1\mbar_1}(z_1) \Phi^{j_2}_{m_2\mbar_2}(z_2)
  & \raisebox{-0.5ex}{$\stackrel{z_1 \to z_2}{\sim}$} & |z_{12}|^{-2h_{12}} 
   \sum_{j_3,m_3,\mbar_3} \ Q(j_a;m_a) \ 
    \Phi^{j_3}_{m_3 \mbar_3}(z_2) \period \label{OPE}
\eqe
$\sum_{j_3,m_3,\mbar_3}$ stands for the summation and integration 
over $j_3,m_3,\mbar_3$, whose precise meaning is not determined yet. 
In our case, $\Phi^{j_1}_{m_1\mbar_1}$ and $\Phi^{j_2}_{m_2\mbar_2}$
belong to $\calD^\pm_j$ or $\calC^\alpha_j$.

Here, one might ask what $\Phi^j_{m\mbar}$ means after the continuation of 
the parameters. However, we will see below that the OPE is almost 
determined by the symmetry and the two- and three-point functions. Thus,
the existence of the correct primary fields is enough in our discussion.  
Namely, 
we do not need the explicit expressions of $\Phi^j_{m\mbar}$, but 
we need only the fact that $\Phi^j_{m\mbar}$  has 
the correct OPE with the $ \widehat{sl}(2)$ currents. 
This is assured by the continuation of the parameters from the $H_3^+$ case, 
which is in accord with the continuation of the correlators. 
Moreover, as mentioned in subsection 2.2, 
the OPE based on such a continuation of $\Phi^j_{m\mbar}$ gives the results 
consistent with the symmetry in the case discussed in 
\cite{Teschner1} and \cite{Hosomichi-S}. It also turns out that this is 
true in our $SL(2,R)$ case. In addition, such a continuation of 
the operators may be supported by considering the analogy to the flat case,
and has been discussed in detail in \cite{Maldacena-O2}.
Classically, $\Phi^j_{m\mbar}$ satisfies the Laplace equation on $SL(2,R)$, 
of course.

Let us return  to (\ref{OPE}). 
$Q$ is some coefficient, and it is determined 
using the two- and three-point functions through
\eqb
   &&  \Bbra \Phi^{j_1}_{m_1\mbar_1}(z_1) \Phi^{j_2}_{m_2\mbar_2}(z_2) 
    \Phi^{j_4}_{m_4\mbar_4}(z_4) \Bket \label{3Q2}  \\ 
  && \qquad  \sim  \ |z_{12}|^{-2h_{12}} 
   \sum_{j_3, m_3, \mbar_3}  Q(j_a;m_a) \ 
    \Bbra \Phi^{j_3}_{m_3 \mbar_3}(z_2) \Phi^{j_4}_{m_4\mbar_4}(z_4)
   \Bket \period \nn
\eqe
There are two possible contributions in the right-hand side from 
$\Bbra \Phi^{j_3}_{m_3 \mbar_3}\Phi^{j_4}_{m_4\mbar_4} \Bket$, which
are proportional to $\delta(j_3-j_4)$ and $\delta(j_3+j_4+1)$, respectively.
Supposed that $\sum_{j_3,m_3,\mbar_3}$ picks up only one of them, 
as is the case in our later discussion, 
the solution to the above condition is given by
\eqb
   Q_\Phi(j_a;m_1,m_2,m_3) &=& \delta^2(m_1+m_2-m_3) W(j_a;m_1,m_2,-m_3)
                              { D(-j_a-1) \o  B_\Phi(j_3,m_3)}
   \period \label{Qjm}
\eqe
This is easily confirmed when the term proportional to $\delta(j_3-j_4)$
is picked up. Even when the other term is picked up, 
the result is the same because of the identity,
\eqb
  Q_\Phi(j_1,j_2,j_3;m_a) B_\Phi(j_3,m_3) & = &  
   Q_\Phi(j_1,j_2,-j_3-1;m_a) A(j_3) 
  \period
\eqe
This is derived by showing that the four terms in (\ref{Wjm2}), 
$W_a$ ($a=1 \sim 4$), satisfy  
\eqb
  {W_a(j_1,j_2,j_3;m_a) \o W_a(j_1,j_2,-j_3-1;m_a)} 
   &=& {D(-j_1-1,-j_2-1,j_3) \o D(-j_1-1,-j_2-1,-j_3-1)} 
   {A(j_3) \o B_{\Phi}(-j_3-1,m_3)}
   \period \label{ratioWa}
\eqe
To see this, we first make use of (\ref{Dj-j-1}), 
so that the right-hand side is simplified to 
\eqb
   \Delta(-j_{13}) \Delta(-j_{23}) 
      {\Gamma(j_3+1+m_3) \Gamma(j_3+1-\mbar_3) 
             \o \Gamma(-j_3+m_3) \Gamma(-j_3-\mbar_3)}
   \period
\eqe
From the second expression of $C^{12}$ in 
(\ref{Cs}) and a similar one for $C^{21}$, 
it follows that $C^{12}$ and $C^{21}$
are invariant under the exchange of $j_3$ and $-j_3-1$ up to factors of the 
gamma functions. It is then straightforward to evaluate the left-hand side 
of (\ref{ratioWa}), and find that each $W_a$ satisfies (\ref{ratioWa}). 
Thus, even if  $\sum_{j_3,m_3,\mbar_3}$ picks 
up both terms in the right-hand side in (\ref{3Q2}), $Q_\Phi$
is a solution up to a factor one half.\footnote{
However, if we assume that $\sum_{j_3,m_3,\mbar_3}$ picks up 
both terms from the beginning, (\ref{3Q2}) gives the condition 
$ Q(j_1,j_2,j_3;m_a) B_\Phi(j_3,m_3) +  
   Q(j_1,j_2,-j_3-1;m_a) A(j_3) 
  = \delta^2(m_1+m_2-m_3) W(j_a;m_1,m_2,-m_3) D(-j_a-1)$. In this case, 
there may be solutions other than $Q_\Phi$.   
}

Let us move on to a discussion about the analytic structure of 
$Q_\Phi$, which is needed later. There are two sources of the poles in $j_3$. 
One is $D(-j_a-1)$. This has simple poles at \cite{Teschner2}
\eqb
   \begin{array}{ll}
            j_3 = \lmb \begin{array}{l}
                          j_1+j_2 +1+ \bfS \comma \\ 
                          j_1+j_2 -b^{-2} - \bfS \comma
                          \end{array} \right.
             & j_3 = \lmb \begin{array}{l}
                         -(j_1+j_2+1) + b^{-2}+\bfS \comma \\
                          -(j_1+j_2+1) -1 - \bfS \comma
                          \end{array} \right.
     \\ 
                j_3 = \lmb \begin{array}{l}
                          j_1 - j_2  + b^{-2} +\bfS \comma \\
                          j_1-j_2 - 1 - \bfS \comma
                           \end{array} \right. 
               & j_3 = \lmb \begin{array}{l}
                         j_2-j_1 +b^{-2} +\bfS \comma \\
                         j_2-j_1-1-\bfS \comma
                       \end{array} \right.
    \end{array}
    \label{poleDja}
\eqe
where $\bfS = \{  l+ nb^{-2} \, \vert \,  l,n \in \bfZ_{\geq 0}  \}$.
The other source is $ W(j_a;m_1,m_2,-m_3) B^{-1}_\Phi(j_3,m_3)$. 
Its analytic structure can be 
studied using the results in appendix C. 
Here, we divide the possible cases 
into the following two: (i) At least one of $\Phi^{j_1}_{m_1\mbar_1}$ 
and $\Phi^{j_2}_{m_2\mbar_2}$ 
is in $\calD^+_j$. (The case of $\calD^-_j$ is similar.)  
(ii) Both $\Phi^{j_1}_{m_1\mbar_1}$ 
and $\Phi^{j_2}_{m_2\mbar_2}$ are in $\calC^\alpha_j$. 

In case (i), we can choose $j_1,m_1,\mbar_1$ so that 
(\ref{lowest}) and, hence, $ W= W_1 $ hold. 
From the discussion in appendix C,  we see 
that the poles and some of the zeros 
in $W_1 B^{-1}_\Phi$  are included in the factor 
\eqb
   && \Delta(-N) \Delta(-j_{12})\Delta(-j_{13})
   {\Gamma(1+j_2+m_2) \o \Gamma(-j_2-\mbar_2)}  
   {\Gamma(-j_3-m_3) \o \Gamma(1+j_3+\mbar_3)} 
  \period
\eqe
Thus, $W_1B^{-1}_\Phi$ has simple poles at 
$N,j_{12},j_{13}, j_3+m_3  \in \bfZ_{\geq 0}$, 
and zeros at $ N,j_{12},j_{13}, j_3+\mbar_3 \in -1 -\bfZ_{\geq 0} $. 
Combining these with the poles in $D(-j_a-1)$, we find the possible poles
in $Q_\Phi$:
\eqb
   \begin{array}{ll}
            j_3 = \lmb \begin{array}{l}
                          j_1+j_2 +1+ b^{-2} +\bfS \comma \\ 
                          j_1+j_2 - \bfS \comma
                          \end{array} \right.
             & j_3 = \lmb \begin{array}{l}
                         -(j_1+j_2+1) +\bfS \comma \\
                          -(j_1+j_2+1) -1 -b^{-2} - \bfS \comma
                          \end{array} \right.
     \\ 
                j_3 = \lmb \begin{array}{l}
                          j_1 - j_2  + b^{-2} +\bfS \comma \\
                          j_1-j_2 - 1 - \bfS \comma
                           \end{array} \right. 
               & j_3 = \lmb \begin{array}{l}
                         j_2-j_1 +\bfS \comma \\
                         j_2-j_1-1-b^{-2} -\bfS \comma
                       \end{array} \right. 
      \\
      j_3 = -m_3 + \bfZ_{\geq 0}  \period &
    \end{array}
    \label{poleD+}
\eqe
These can collide with each other. 
In such a case, we regularize the parameters so that they remain 
simple poles.

In case (ii), it is useful to go back to (\ref{Wjm1}),
and use the expressions in (\ref{Cs}) for 
 $C^{31}$, $C^{12}$ (the second one), 
and similar ones for $C^{21}$ and $C^{13}$.
Then, from the results in appendix C, the poles in $W$
are read off from some factors of the gamma functions. Collecting the relevant
factors, we see that the poles in $WB^{-1}_\Phi$ are included in 
\eqb
  \Gamma(-N) \Gamma^2(-j_{12}) \Gamma(-j_{13}) 
  \Gamma(-j_3-m_3) \Gamma(-j_3+\mbar_3)\Gamma(1+j_3+m_3)
  \comma
\eqe
and a similar one with $ j_1$ and $j_2$ exchanged.
Combining these poles with those in (\ref{poleDja}), we find 
the possible poles in $Q_\Phi$ in this case:
\eqb
     \begin{array}{ll}
            j_3 = \lmb \begin{array}{l}
                          j_1+j_2 +1+ \bfS \comma \\ 
                          j_1+j_2 - \bfS^\ast \comma
                          \end{array} \right.
             & j_3 = \lmb \begin{array}{l}
                         -(j_1+j_2+1) +\bfS \comma \\
                          -(j_1+j_2+1) -1 - \bfS \comma
                          \end{array} \right.
     \\ 
                j_3 = \lmb \begin{array}{l}
                          j_1 - j_2  +\bfS \comma \\
                          j_1-j_2 - 1 - \bfS \comma
                           \end{array} \right. 
               & j_3 = \lmb \begin{array}{l}
                         j_2-j_1 +\bfS \comma \\
                         j_2-j_1-1-\bfS \comma
                       \end{array} \right.
      \\
              j_3 = -m_3+ \bfZ_{\geq 0} \comma 
                 & j_3 = -m_3 - 1-\bfZ_{\geq 0}
       \\
             j_3 = \mbar_3 + \bfZ_{\geq 0} \period &
    \end{array}
    \label{poleCC}
\eqe
Here, $\ast $ indicates that that sequence includes possibly double poles
coming from $\Gamma^2(-j_{12})$. (It may be possible that several terms 
sum up to make them simple poles.) The list of the 
$(m_3,\mbar_3)$-independent poles is the same as the one appeared in the
OPE of $\Phi_j$ \cite{Teschner2} up to $\ast$. 

\mysubsection{Classical limit}
The precise meaning of $\sum_{j_3,m_3,\mbar_3}$ is yet to be determined. 
For this purpose, we consider the classical limit in this 
subsection. This is along the line of the arguments which determined the 
OPE in the $H_3^+$ WZW model: In \cite{Teschner1,Teschner2}, 
the OPE in the $H_3^+$ WZW model was proposed by using several 
arguments, one of which was a consideration of the classical limit.
The crossing symmetry of the four-point functions has been shown by using 
the proposed OPE in \cite{Teschner3}. 

In the classical limit $ k \to \infty$ (or particle limit), 
the problem reduces to that of zero-modes. 
The OPE of $\Phi^j_{m\mbar}$ 
then should represent the decomposition of a product of two
`wave functions' $\Phi^{j_1}_{m_1\mbar_1} \times \Phi^{j_2}_{m_2\mbar_2}$ 
into a set of  `wave functions' $\{ \Phi^{j_3}_{m_3 \mbar_3} \}$.
Since $\Phi^{j_1}_{m_1\mbar_1}, \Phi^{j_2}_{m_2\mbar_2}$ belong
to the normalizable unitary representations of $SL(2,R)$, 
they correspond to the elements of the $L^2$-space on $SL(2,R)$.
According to the harmonic analysis, 
$\Phi^{j_1}_{m_1\mbar_1} \times  \Phi^{j_2}_{m_2\mbar_2}$ also corresponds to 
an element of the $L^2$-space and, hence, 
should be decomposed by the normalizable unitary representations.

Therefore, (at least) in this limit, $\sum_{j_3,m_3,\mbar_3}$
should be the summation over $\calD^\pm_j$ and $\calC^\alpha_j$, 
namely,
\eqb
   \sum_{j_3,m_3,\mbar_3} & = &
     \int_{-\infty}^\infty dm_3 \int_{-\infty}^\infty d\mbar_3
    \lbb \int_{\calP_+} dj_3 + \delta_{\calD^\pm_j} \oint_{\calC} dj_3 \rbb
   \period \label{sumsl2}
\eqe
Some explanations may be in order here: 
Together with $\delta^2(m_1+m_2-m_3)$,  
the integrations over $m_3$ and $\mbar_3$
give the conservation of $m,\mbar$. The integration over  
$\calP_+ = -1/2 + i \bfR_{> 0}$ stands for the 
summation over $\calC^\alpha_j$. The summation over $\alpha$ is 
already encoded in that of $m,\mbar$.
The last contour integral indicates the summation 
over $\calD^\pm_j$. $\delta_{\calD^\pm_j}$ means that 
$j_3$ is picked up only when $ \{ j_3,m_3,\mbar_3 \} $ are  
the quantum numbers of $\calD^\pm_j$. In such a case, 
the corresponding $j_3$ comes from the pole in $Q_\Phi$ picked up by 
the contour $\calC$. Note that the range of $j_3$ is Re$ \, j_3 \leq -1/2$ 
and Im$ \, j_3 \geq 0$. This is consistent with the argument which determined 
$Q_\Phi$, because $\sum_{j_3,m_3,\mbar_3}$ picks up only one term in the
right-hand side in (\ref{3Q2}).

In addition, in the limit $k \to \infty$,
the possible poles in $Q_\Phi$ become
 \eqb
   \begin{array}{ll}
            j_3 = \lmb \begin{array}{l}
                          -(j_1+j_2+1) +\bfZ_{\geq 0} \comma  \\
                         j_1+j_2 - \bfZ_{\geq 0} \comma
                          \end{array} \right.
  &
            j_3 = \lmb \begin{array}{l}
                         j_2-j_1 +\bfZ_{\geq 0} \comma \\
                          j_1-j_2 - 1 - \bfZ_{\geq 0} \comma
                           \end{array} \right. 
      \\
      j_3 = -m_3 + \bfZ_{\geq 0}  \period &
    \end{array}
    \label{ClpoleD+}
\eqe
in the case where $\Phi^{j_1}_{m_1,\mbar_1}$ belongs to $\calD^+_j$.
In the case where both
$\Phi^{j_1}_{m_1\mbar_1}$ and $\Phi^{j_2}_{m_2 \mbar_2}$ belong to 
$ \calC^\alpha_j $, the possible poles are obtained 
simply by replacing $\bfS$ in (\ref{poleCC}) with $\bfZ_{\geq 0}$.

In order to further determine the contour $\calC$, 
we need to study  which poles are compatible with the quantum 
numbers of $\calD^\pm_j$. As an example, let us consider the 
case where $\Phi^{j_1}_{m_1\mbar_1}, \Phi^{j_2}_{m_2\mbar_2} 
\in \calD^+_j \otimes \calD^+_j$, namely, $ m_{1,2}= -j_{1,2} + n_{1,2}$ 
($n_{1,2} \in \bfZ_{\geq 0}$) (and similarly for $\mbar_{1,2}$). 
The conservation of $m,\mbar$ then implies that 
$m_3 = m_1 + m_2 = -j_1-j_2 +n_1 +n_2$. Thus, for 
$\Phi^{j_3}_{m_3,\mbar_3} \in \calD^+_j \otimes \calD^+_j $ with 
$ j_3 = -m_3 + n_3 $ ($n_3 \in \bfZ_{\geq 0}$), only 
$ j_3 = -m_3 + \bfZ_{\geq 0}$ and a part of 
$ j_3 = j_1+j_2 - \bfZ_{\geq 0}$ in 
(\ref{ClpoleD+}) can satisfy the constraint $\delta_{\calD^\pm_j}$. 
Similarly, for $\Phi^{j_3}_{m_3,\mbar_3} \in \calD^-_j \otimes \calD^-_j $,
the possible sequence is $j_3 = -(j_1+j_2+1) + \bfZ_{\geq 0} $. However,
in this case, the residues vanish because of the factor 
$ \Gamma^{-1}(-j_2 -\mbar_2)$ in $Q_\Phi$.
In this way, we obtain a table of the allowed poles (Table 1).

From this table, we notice that there are two allowed sequences in 
(1a) and (2a). In (1a), 
$j_3 = -m_3 + n_3 = j_1+j_2 -n_1 -n_2 + n_3$. Hence, a part of 
the two sequences degenerates. We regularize such degeneracy by slightly
shifting the parameters. The residues are then non-vanishing
only for such degenerating poles because of the factor 
$ \Gamma^{-1}(-j_2-\mbar_2)$ in $Q_\Phi$. To avoid double counting, 
we take the contour $\calC$ so that it encloses the poles 
in the sequence $j_3 = -m_3 + \bfZ_{\geq 0}$. Consequently, 
$j_3= j_1+j_2, j_1+j_2-1, \cdots $ contribute to the OPE. 

In (2a), $m_1 = -j_1 + n_1$, $m_2 = j_2-n_2$ and 
$j_3 = -m_3 + n_3 = j_1-j_2 - n_1+n_2 +n_3$. Thus, it 
is possible again that a part of the two sequences degenerates. However,  
in such a case, the corresponding residues are divergent 
(assuming the regularization of the degeneracy). 
Since $Q_\Phi$ is essentially a three-point function, 
this may indicate that the three-point function is ill-defined. 
Therefore, we take $\calC$ so that it encloses the poles in 
the sequence $j_3 = -m_3 + \bfZ_{\geq 0}$ whose residues are regular.
Consequently, $j_3 = j_1-j_2, j_1-j_2+1, \cdots < -1/2 $ contribute
to the OPE. 

For other cases, we take $\calC$ so that it simply encloses 
the poles in Table 1. In sum,
the contour $\calC$ encloses the poles,  
$ j_3 = -m_3+ \bfZ_{\geq 0}, \ \mbar_3 + \bfZ_{\geq 0}, \ 
j_2-j_1 + \bfZ_{\geq 0} \, < -1/2 $, whose residues 
are not divergent.

%
%\par\bigskip
\newpage
\newcommand{\lw}[1]{\smash{\lower2.0ex\hbox{#1}}}
\begin{center}
\begin{tabular}{|c|c|c|c|l|}
  \hline
  $\Phi^{j_1}_{m_1\mbar_1}$ & $\Phi^{j_2}_{m_2\mbar_2}$ & 
     $\Phi^{j_3}_{m_3\mbar_3}$
   &  
   & \hspace{2.7em} {\rm allowed poles } $(j_3 < -1/2)$ \\
  \hline 
  \lw{$\calD^+_j$} & \lw{$\calD^+_j$} & $\calD^+_j$ & (1a) &
     $ j_3 = -m_3 + \bfZ_{\geq 0} \comma $ \  
     $ j_3 = j_1+j_2 - \bfZ_{\geq 0} $  \\
  \cline{3-5}
  & &  $\calD^-_j$ & (1b) & \hspace{7em} $-$ \\
  \hline
  \lw{$\calD^+_j$} & \lw{$\calD^-_j$} & $\calD^+_j$ & (2a) & 
     $ j_3 = -m_3 + \bfZ_{\geq 0} \comma $ \  
     $ j_3 = j_1 - j_2 - 1- \bfZ_{\geq 0} $  \\
  \cline{3-5}
  & &  $\calD^-_j$ & (2b) & $ j_3 = j_2-j_1 + \bfZ_{\geq 0}$ \\
  \hline
  \lw{$\calD^+_j$} & \lw{$\calC^\alpha_j$} & $\calD^+_j$ & (3a) &
     $ j_3 = -m_3 + \bfZ_{\geq 0}$ \\
  \cline{3-5}
  & &  $\calD^-_j$ & (3b) & \hspace{7em}  $-$  \\
  \hline
  \lw{$\calC^\alpha_j$} & \lw{$\calC^\alpha_j$} & $\calD^+_j$ & (4a) &
     $ j_3 = -m_3 + \bfZ_{\geq 0}$ \\
  \cline{3-5}
  & &  $\calD^-_j$ & (4b) & $ j_3 = \mbar_3 + \bfZ_{\geq 0}$  \\
  \hline
\end{tabular}
\end{center}
\par\ni
\baselineskip = 0.5cm
{\small {\bf Table 1 :} \
The first three columns show the representations of 
$\Phi^{j_a}_{m_a\mbar_a}$ ($a = 1,2,3$), respectively. We have used the
abbreviations $\calD^+_j$ for $\calD^+_j \otimes \calD^+_j$, and so on.
For the $(m_3,\mbar_3)$-independent poles, only 
a part of the sequences satisfying the conservation of $m,\mbar$ 
is allowed. }

\baselineskip = 0.6cm
\vspace{3ex}

Here, we would like to make two remarks. One is about case (2b).
In this case,  
$ j_3 =j_2 -j_1 + \bfZ_{\geq 0} = m_3 - (n_1 -n_2) + \bfZ_{\geq 0}$.
Thus, $ j_3 < m_3 $ seems to appear for $ n_1 > n_2$, which 
is out of the range of $\calD^{-}_{j}$. However, we can show that 
the residues in such cases automatically vanish: First, 
for $ j_3 = m_3 -1$, the residue vanishes because of (\ref{SaalschutzF}).
Furthermore, using an identity similar to (\ref{ladder}), 
we see that the residues for $j_3 -m_3 < -1$ also vanish.

The other remark is about the selection rules from the right sector.
In the above, we have concentrated on the left sector. However, 
since the left and the right sectors are symmetric, so should be 
the OPE. In particular, $\Phi^{j_3}_{m_3\mbar_3}$ with 
$j_3 \pm \mbar_3 \in -1 -\bfZ_{\geq 0}$ 
should not appear. In (1a),(2a) and (3a) where 
$ W = W_1$ and $ j_3 +m_3 \in \bfZ_{\geq 0}$, this is accounted by the factor 
$ \Gamma^{-1}(1+j_3+\mbar_3)$ in $Q_\Phi$. 
As a consequence, the contributions, for example, in (1a)
come from $j_3= j_1+j_2, j_1+j_2-1, \cdots, j_1+j_2 -n$ 
with $n =$ min$\{n_1+n_2, \nbar_1+\nbar_2 \}$. In addition, 
the zeros from $ \Gamma^{-1}(1+j_3+\mbar_3) $ can cancel the divergence 
in the above discussion about (2a). 
Thus, precisely speaking, we should have said above that $\calC$ picks up 
the poles in the sequence $j_3 = -m_3 + \bfZ_{\geq 0}$ 
whose residues are regular for generic $\nbar_1, \nbar_2$.
 
In (2b) where $ W = W_1$ and $ j_3 -m_3 \in \bfZ_{\geq 0}$, 
the mechanism in the above remark about (2b) works also for 
$(j_3,\mbar_3)$. This ensures the correct truncation. For the last two
cases in Table 1, we may need to carry out the summation of the terms in $W$
carefully in order to check the absence of the unwanted contributions.
However, this is indirectly confirmed by the fact that 
we can repeat the analysis in this section with 
$m_a$ and $\mbar_a$ exchanged.     

Now that the meaning of $\sum_{j_3,m_3,\mbar_3}$ has been determined, 
the OPE is obtained by 
collecting the contributions from $\calP_+$ and $\calC$.
Since we can exchange the roles of $\calD^+_j$ and $\calD^-_j$  
by redefining $(J^\pm_0,J^3_0)$ as $(J^\mp_0,-J^3_0)$, 
the case with $\calD^+_j$ and $\calD^-_j$ exchanged is similar.
Thus, we find the following results in the $k \to \infty$ limit:
\eqb
  \Bigl[ {\calD}^{\pm}_{j_1} \otimes {\calD}^{\pm}_{j_1} \Bigr]
    \, \otimes \, 
  \Bigl[ {\calD}^{\pm}_{j_2} \otimes {\calD}^{\pm}_{j_2} \Bigr]
    & \to &  \sum_{j_3 \leq j_1+j_2} 
      \Bigl[ {\calD}^{\pm}_{j_3} \otimes {\calD}^{\pm}_{j_3} \Bigr]
    \comma 
    \label{clope1} \\
  \Bigl[ {\calD}^{+}_{j_1} \otimes {\calD}^{+}_{j_1} \Bigr]
    \, \otimes \, 
   \Bigl[ {\calD}^{-}_{j_2} \otimes {\calD}^{-}_{j_2} \Bigr]
     & \to &
     \int_{\calP_+}dj_3 \, 
   \Bigl[ {\calC}^{\alpha_3}_{j_3}  \otimes 
        {\calC}^{\alpha_3}_{j_3}  \Bigr]
    \label{clope2} \\
   && \quad  \oplus 
      \sum_{j_1-j_2 \leq j_3 < -\half} 
   \Bigr[ {\calD}^{+}_{j_3} \otimes {\calD}^{+}_{j_3} \Bigr]   
      \oplus 
      \sum_{j_2-j_1 \leq j_3 < -\half} 
   \Bigr[ {\calD}^{-}_j \otimes {\calD}^{-}_j \Bigr]  
   \comma   \nn  \\
  \Bigl[ {\calD}^{\pm}_{j_1} \otimes {\calD}^{\pm}_{j_1} \Bigr]
    \, \otimes \, 
   \Bigl[ {\calC}^{\alpha_2}_{j_2} \otimes 
          {\calC}^{\alpha_2}_{j_2}  \Bigr]
  & \to & 
    \int_{\calP_+} dj_3 \, 
    \Bigl[ {\calC}^{\alpha_3}_{j_3} \otimes 
            {\calC}^{\alpha_3}_{j_3} \Bigr] 
   \oplus 
   \sum_{ j_3 < -\half} 
   \Bigr[ {\calD}^{\pm}_{j_3} \otimes {\calD}^{\pm}_{j_3} \Bigr]   
   \comma \label{clope3} \\
  \Bigl[ {\calC}^{\alpha_1}_{j_1}  \otimes 
                {\calC}^{\alpha_1}_{j_1}  \Bigr]
   \, \otimes \, 
   \Bigl[ {\calC}^{\alpha_2}_{j_2}  \otimes 
               {\calC}^{\alpha_2}_{j_2}  \Bigr]
  & \to & 
    \int_{\calP_+} dj_3 \, 
    \Bigl[ {\calC}^{\alpha_3}_{j_3}  \otimes 
                {\calC}^{\alpha_3}_{j_3}  \Bigr] 
   \oplus 
   \sum_{ j_3 < -\half } 
   \Bigr[ {\calD}^{+}_{j_3} \otimes {\calD}^{+}_{j_3} \Bigr]
    \oplus \sum_{ j_3 < -\half} 
   \Bigr[ {\calD}^{-}_{j_3} \otimes {\calD}^{-}_{j_3} \Bigr]   
     \period \nn \\
  && \label{clope4} 
\eqe
In the above, $j_3$ for the discrete series is summed with integer spacing.
$\alpha_3$ for the continuous series and $j_3$ for the discrete series 
take the values so that they are compatible with the conservation
of $m,\mbar$. 
Note that the contributions from the continuous series are absent
in (\ref{clope1}) because the integrand $ Q_\Phi $ vanishes.
In (\ref{clope4}), we have not cared about 
the multiplicity of the representations in the right-hand side.
For the tensor products of the $SL(2,R)$ representations,
the continuous series appears twice in the corresponding case.
This is due to the existence of two linearly independent 
Clebsch-Gordan coefficients.
In our case, such multiplicity corresponds 
to the fact that both $C^{12}$ and $C^{21}$
contribute to the OPE. These results are in complete agreement 
with the tensor products of the $SL(2,R)$ 
representations, which are given in (\ref{sl2tensor}).\footnote{
Precisely, we need to confirm that the residues of the contributions to 
(\ref{clope1})-(\ref{clope4}) are non-vanishing. However, we 
do not see any special reason for them to vanish.
}

\mysubsection{Relation to \cite{Hosomichi-S}}
The OPE in the $SL(2,R)$ WZW model was also discussed 
in \cite{Hosomichi-S} for finite $k$, using the three-point functions
with $j_1 +m_1 = j_1 + \mbar_1 = 0$.
In this subsection, we would like to discuss the relation between 
the arguments in the previous subsection and  
in \cite{Hosomichi-S} (after taking the limit $k \to \infty$).

First, the primary fields used in \cite{Hosomichi-S} were
\eqb
  V^{j}_{m\mbar} & = & \gamma^{j+m}\gammabar^{j+\mbar} e^{2j\phi}
   \comma
\eqe
where $\gamma,\gammabar, \phi$ are certain  coordinates of $H_3^+$ or
$SL(2,R)$. These are nothing but the primary fields in the standard Wakimoto
realization of $\widehat{sl}(2) \times \widehat{sl}(2)$. 
Since $\Phi_j = (|\gamma-x|^2 e^{\phi} + e^{-\phi})^{2j}$ 
in terms of $\gamma,\gammabar, \phi$, 
the relation between $\Phi^{j}_{m\mbar}$ and $V^{j}_{m\mbar}$ 
is given by
\eqb
  \Phi^{j}_{m\mbar} \Big{\vert}_{\phi \to \infty} &=& 
   c^{-j-1}_{m\mbar} V^{-j-1}_{m\mbar} 
  \period \label{PhiV} 
\eqe
Here, $\phi \to \infty $ is the `free-field' 
limit (see, e.g., \cite{Ishibashi-OS,Hosomichi-OS}). Note, however, that
we do not need  such a limit in the discussion using $\Phi^{j}_{m\mbar}$.
Because of the intertwiner $c^{j}_{m\mbar}$, 
$V^{j}_{m\mbar}$ form a representation with spin $j$ (not $-j-1$) 
as $\Phi^{j}_{m\mbar}$.

The correlation functions of $V^{j}_{m\mbar}$ are obtained by 
substituting the right-hand side of (\ref{PhiV}) into those of 
$\Phi^j_{m\mbar}$, namely, (\ref{Phijm2pt}) and (\ref{Phijm3pt}). 
This is shown by direct calculations \cite{Hosomichi-S}. 
Then, the OPE of $V^{j}_{m\mbar}$ can be discussed similarly to 
the case of $\Phi^j_{m\mbar}$.  In this case, the quantity
corresponding to $Q_\Phi$ turns out to be
\eqb
  Q_V(j_a;m_a) &=&   {R(j_1)R(j_2) \o R(j_3)} Q_\Phi(j_a;m_a)
  \comma 
\eqe
with $R(j) = B(j)/A(j)$. To derive this relation, we have used 
(\ref{ratioWa}). $Q_V$ reduces to the expression in \cite{Hosomichi-S} when
$j_1+m_1 = j_1+\mbar_1 = 0$. 
Since $R(j)$ has no poles and zeros for generic 
$j$ and $k$, the discussions of the OPE are essentially 
the same for $\Phi^j_{m\mbar}$ and $V^{j}_{m\mbar}$.   

Second, the main idea in \cite{Hosomichi-S} was that 
the OPE in the $SL(2,R)$ WZW model is obtained from 
the OPE in the $H_3^+$ WZW model by continuing the 
parameters appropriately. This is based on the arguments 
in \cite{Teschner1,Teschner2}, which proposed that the OPE in 
a model with an $\widehat{sl}(2)$ symmetry is obtained by such a continuation.
With such a prescription, the OPE includes contributions from the continuous
series by definition (unless the coefficients
vanish), because the OPE in 
the $H_3^+$ WZW model is of the form 
$\Phi_{j_1} \Phi_{j_2} \sim \int_{\calP_+} dj_3 \Phi_{j_3}$. 
In \cite{Hosomichi-S}, it was further argued that  
the OPE in the $SL(2,R)$ WZW model 
has other contributions corresponding to case (2b) in 
Table 1 by continuing $j_1$ and $j_2$ from 
the values of the continuous series.  However, 
how to precisely deal with the $(m_3,\mbar_3)$-dependent poles, 
which are absent in the discussion of the $H_3^+$ WZW model, was an 
open question.

We have taken an apparently different strategy in this section. 
Nevertheless, for the cases covered in \cite{Hosomichi-S} we notice that 
our results are obtained by further picking up 
the appropriate $(m_3,\mbar_3)$-dependent poles in \cite{Hosomichi-S}.
Thus, after all our prescription is regarded 
as a completion of the arguments in \cite{Hosomichi-S}. 
This means that the arguments in \cite{Hosomichi-S}
and in this paper support each other.
We remark that the issue of the divergence in case (2b) did not appear in 
\cite{Hosomichi-S}, because $n_1 = \nbar_1 = 0$ there. 

\mysubsection{Full OPE}
Now we would like to consider the full OPE for finite $k$. 
First, let us summarize our proposal for the OPE.
Taking into account the discussion in the previous subsection,
it can be stated as follows:

\begin{description}
 \item[\rm (1)] The OPE is given by (\ref{OPE}) with $Q = Q_\Phi$ 
         in (\ref{Qjm}) and $\sum_{j_a,m_3,\mbar_3}$ in (\ref{sumsl2}).
 \item[\rm (2)] The contour $\calC$ picks up the following poles in 
     the region $ j_3 <-1/2$: (i) the $(m_3,\mbar_3)$-dependent
            poles of the type $j_3 = \pm m_3 + \bfZ_{\geq 0}$, 
         (ii) the poles 
           which cross over $\calP_+$ when $j_1,j_2$ are supposed to be  
            continued from $j_1,j_2 \in \calP_+$.
 \item[\rm (3)] A pole whose residue is (generically) divergent 
         should not be picked up.  
\end{description}
For instance, $j_3 = j_2 -j_1 + \bfZ_{\geq 0}$ are on the right-hand side 
of $\calP_+$ in the complex $j_3$-plane when $j_1,j_2 \in \calP_+$. 
As  $j_1,j_2$ are continued to $j_1,j_2 < -1/2$, some of $j_3$'s above 
cross over $\calP_+$, and such poles are picked up. (For more details,
see \cite{Teschner1,Teschner2,Hosomichi-S}.) 

Although this prescription was given for $ k \to \infty$, it can be 
applied to the case of finite $k$ without modifications. 
Thus, we take the above (1)-(3) as the definition of the full OPE.
The analysis according to this definition is then straightforward. 
As a consequence, it turns out that all the additional poles appearing for 
generic finite $k$
do not contribute because of the conservation of $ m,\mbar$. 
Therefore, we arrive at the results which are essentially the same
as those in the $k \to \infty$ limit. 
They are given just by replacing ${\calD}^\pm_j, {\calC}^\alpha_j$
in (\ref{clope1})-(\ref{clope4}) with the corresponding affine representations 
$\hat{\calD}^\pm_j, \hat{\calC}^\alpha_j$.

Those results 
show that the OPE of the primary fields in the $SL(2,R)$ WZW model
is essentially the same as the classical tensor products.
Here, it would be useful to recall the case of the 
$SU(2)_{\tilde{k}}$ WZW model. 
In that case, the existence of the non-trivial null states dictates 
the decoupling of the primary fields with $SU(2)$ spin 
$ \tilde{j} > \tilde{k}/2$. This mechanism makes the 
difference from the classical tensor products \cite{Gepner-W}. However,
in the case of $SL(2,R)$, there are no corresponding 
non-trivial null states. Thus, it is natural that 
the full OPE (\ref{clope1})-(\ref{clope4}) with 
$\hat{\calD}^\pm_j, \hat{\calC}^\alpha_j$
is essentially the same as the classical tensor products (\ref{sl2tensor}). 
Note that any correct OPE should recover (\ref{sl2tensor}) 
in the limit $k \to \infty$.

\mysection{Discussion}
In this paper, we first wrote down the explicit expression of 
the three-point functions of $\Phi^j_{m\mbar}$, and analyzed its
properties. Based on these results, we discussed the OPE in the
$SL(2,R)$ WZW model. Guided by a consideration in the classical 
(particle) limit, we proposed the full OPE. It was essentially the same as
the classical tensor products of the representations of  
$SL(2,R)$. This is natural 
because we may not have any reason that makes difference, contrary to the 
$SU(2)$ case. We also noted that our prescription 
is regarded as a completion of the arguments in 
\cite{Hosomichi-S}, though our strategy was apparently different.
For a further check of the validity of our OPE, it is desirable to 
show the crossing symmetry of the four-point functions by using the OPE, 
as has been done for the $H_3^+$ WZW model \cite{Teschner3}.  

A main problem in the $SL(2,R)$ WZW model
has been to determine the correct spectrum. 
A recent proposal in \cite{Maldacena-O} requires that  
the spin $j$ in the discrete series be in the `unitarity bound' 
$ -(k-1)/2 < j < -1/2$.
Having a look at the OPE in (\ref{clope1})-(\ref{clope4}) with 
$\hat{\calD}^\pm_j, \hat{\calC}^\alpha_j$, we then find a puzzle:
The OPE of the type (\ref{clope1}) seems to break 
the unitarity bound. We remark that such a puzzle exits independently of  
our prescription, because any correct OPE 
should include that kind of contributions at least for sufficiently large $k$.
Possibilities of the resolution might be, for example, as follows. 
(i) As in the case of the partition function \cite{Kato-S}, 
    there may be different expressions for the same quantity 
     (up to some formal manipulations), because there are infinitely many
   primary fields. Thus, one may be able to 
   rewrite the OPE of the type (\ref{clope1})
   in a form compatible with the unitarity bound. The spectral-flowed 
    sectors may appear in such a process. 
    A similar mechanism is discussed about $\Phi_j$ 
   in the $k \to \infty$ limit \cite{Hosomichi-S}.
(ii) In the string theory context, the $SL(2,R)$ WZW model appears 
    typically for describing the strings on 
    $AdS_3 \times S^3 \times {\cal M}$.
   In such a case, a primary field of the model 
    is a tensor product of the three parts. For the $SU(2)$ part,  
    there is a bound on the $SU(2)$ spin.  
  Since the $SU(2)$ and $SL(2,R)$ spins are related by 
   the physical state conditions, the bound on the $SU(2)$ spin may 
  induce also the bound on the $SL(2,R)$ spin.
   A related discussion is found in \cite{Petropoulos-R}. 
   Thus, the unitarity bound may be maintained in the case of the string 
    theory on $AdS_3 \times S^3 \times {\cal M}$.
    There is a possibility that we might be missing something important.    
   In any case, it seems to be important to investigate this problem further.

\vspace{4ex}
\begin{center}
  {\bf Acknowledgments}
\end{center}
  The author would like to thank T. Eguchi, N. Ishibashi, K. Mohri 
  and, especially, K. Hosomichi for useful discussions and conversations.
  This work was supported in part  
 by Grant-in-Aid for Scientific Research on Priority Area 707 and
 Grant-in-Aid for Scientific Research (No.12740134) from
 the Japan Ministry of Education, Culture, Sports, Science and Technology. 
 
  I would like to dedicate this paper to the memory 
  of Professor Sung-Kil Yang.

\vspace{1ex}
\setcounter{section}{0}
\appsection{Useful formulas}
In appendix A, we collect the formulas used in the main text.

\ni (1) \
 In \cite{Fukuda-H}, the following integral
   has been carried out: 
\eqb
  I &=& \int d^2z d^2w \, z^\alpha(1-z)^{\beta} 
        \zbar^{\alphabar}(1-\zbar)^{\betabar} w^{\alpha'}(1-w)^{\beta'}
        \wbar^{\alphabar'}(1-w)^{\betabar'} |z-w|^{4\sigma} \nn \\
    &=& (i/2)^2 
   \lmb C^{12}[\alpha_i,\alpha'_i] P^{12} [\alphabar_i,\alphabar'_i] 
    + C^{21}[\alpha_i,\alpha'_i] P^{21} [\alphabar_i,\alphabar'_i] \rmb
  \period \label{I1}
\eqe
Here, 
\eqb
   C^{ab}[\alpha_i,\alpha'_i] &=& 
    {\Gamma(1+\alpha_a+\alpha'_a-k')\Gamma(1+\alpha_b+\alpha'_b-k')
   \o \Gamma(k'-\alpha_c-\alpha'_c)} 
   \G{\alpha'_a+1,\alpha_b+1,k'-\alpha_c-\alpha'_c}{\alpha'_a-\alpha_c+1,
     \alpha_b-\alpha'_c+1} \comma \nn \\
   \G{a,b,c}{e,f} &=& {\Gamma(a)\Gamma(b)\Gamma(c) \o \Gamma(e)\Gamma(f)}
         \F{a,b,c}{e,f} \comma \qquad 
    \F{a,b,c}{e,f} \ = \ \3F2 (a,b,c;e,f;1) \comma 
    \label{Cab}
\eqe
\eqb
  && \alpha_1 = \alpha \comma \, \quad \alpha_2 = \beta \comma \, \quad 
     \alpha_3 = \gamma \comma \quad \ \alpha + \beta + \gamma + 1 = k' 
  = -2\sigma -1 \comma \nn \\
  && \alpha'_1 = \alpha' \comma \quad \alpha'_2 = \beta' \comma \quad 
     \alpha'_3 = \gamma' \comma \quad \alpha' + \beta' + \gamma' + 1 = k' 
  = -2\sigma -1 \comma \nn
\eqe
and similarly for $\alphabar_i$ and $\alphabar'_i$.
($\gamma,\gamma',\gammabar, \gammabar'$ are determined through
the above type of equations.) 
$P^{12}$ and $P^{21}$ are expressed by $C^{ab}$ as
\eqb
  && (i/2)^2 {P^{12} \brack P^{21}}
     = A_\beta {C^{23} \brack C^{32} }
     = A^T_\alpha { C^{31} \brack C^{13} }
   \comma \quad  
    A_\alpha = \Biggl[ \begin{array}{cc} 
   s(\alpha)s(\alpha') &  -s(\alpha) s(\alpha'-k') \\ 
     -s(\alpha')s(\alpha-k') & s(\alpha)s(\alpha') \end{array} \Biggr] 
  \comma \nn \\
  && \label{P12}
\eqe
with $s(x) = \sin (\pi x)$.
Note that $C^{ba}$ and, hence, $P^{ba}$ are obtained by exchanging
$\alpha_i $ and $\alpha'_i$ in $C^{ab}$ and $P^{ab}$. Namely,
\eqb 
 C^{ba} &=& C^{ab}(\alpha_i \leftrightarrow \alpha'_i) \comma \quad 
  P^{ba} \ = \ P^{ab}(\alpha_i \leftrightarrow \alpha'_i)
\period
 \label{CbaCab}
\eqe
\par\medskip\ni
(2) \ 
  Among $ \F{a,b,c}{e,f} $ or $ \G{a,b,c}{e,f} $, many relations
  are known \cite{Slater}. For example, 
\eqb
  \G{a,b,c}{e,f} &=& {\Gamma(b)\Gamma(c) 
    \o \Gamma(e-a) \Gamma(f-a)} \G{e-a,f-a,u}{u+b,u+c}
   \label{2G} \\ 
   &=& {\Gamma(b)\Gamma(c) \Gamma(u) \o \Gamma(f-a) \Gamma(e-b)\Gamma(e-c)} 
  \G{a,e-b,e-c}{e,a+u} \comma \nn  \\
  \G{a,b,c}{e,f} &=& {s(e-b)s(f-b) \o s(a)s(c-b)} 
   \G{b,1+b-e,1+b-f}{1+b-c,1+b-a} \label{3G}   \\
    && \qquad \quad  + \ 
  {s(e-c)s(f-c) \o s(a)s(b-c)} 
   \G{c,1+c-e,1+c-f}{1+c-b,1+c-a} 
  \comma \nn 
\eqe
where $u = e+f-a-b-c$. The two-term relations in (\ref{2G})
generate identities among ten apparently different expressions. 
As discussed in appendix C, 
we need the value of $u$ to study the analytic
structure of $F$ and $G$. 
It changes as $u \to a \to f-a$ in (\ref{2G}), whereas $u$ is invariant
in (\ref{3G}). (If we substitute 
the parameters in (\ref{intWjm}) into $\alpha_i,\alpha'_i$ in (\ref{Cab}), 
$u=-j_{12}$ for any $C^{ab}[\alpha_i,\alpha'_i]$.)

\par\medskip\ni
(3) \ From the properties of $\Upsilon(x)$, it follows that
\eqb
    D(j_1,j_2,j_3) & = & \pi \Delta(-j_{13}) \Delta(-j_{23})
     \Delta(2j_3+1) R(j_3) D(j_1,j_2,-j_3-1) 
  \comma \label{Dj-j-1}
\eqe
with $R(j)= B(j)/A(j)$.
Note that $D(j_a)$ are symmetric with respect 
to $j_1,j_2,j_3$. 
\par\medskip\ni
(4) \ The tensor products of the normalizable unitary representations
   of $SL(2,R)$ are given by \cite{Vilenkin-K,Holman-B,Wang}
\eqb
    \calD^{\pm}_{j_1} \otimes \calD^{\pm}_{j_2}
    &=& \sum_{j\leq j_1+j_2} \oplus \ \calD^{\pm}_{j} \comma
   \nn \\ 
  \calD^{+}_{j_1} \otimes \calD^{-}_{j_2}
    &=& \int_{\calP_+}dj \, \calC^\alpha_j \, \oplus 
   \sum_{j_1-j_2 \leq j < -\half} \calD^{+}_{j} 
   \, \oplus  \sum_{j_2-j_1 \leq j < -\half} \calD^{-}_{j}  
    \nn   \\
  \calD^{\pm}_{j_1} \otimes \calC^{\alpha_2}_{j_2}
      &=& \int_{\calP_+}dj \, \calC^\alpha_j \, \oplus 
           \sum_{j < -\half} \calD^{\pm}_j \comma 
           \label{sl2tensor} \\
   \calC^{\alpha_1}_{j_1} \otimes \calC^{\alpha_2}_{j_2}
    &=& \int_{\calP_+}dj \, \calC^\alpha_j \, \oplus
        \int_{\calP_+}dj' \, \calC^{\alpha'}_{j'} \, \oplus
   \sum_{j< -\half} \lb \calD^{+}_j \oplus \calD^{-}_j \rb
    \nn \comma
\eqe 
with $\calP_+ = -1/2 + i \bfR_{> 0}$.
Here, $j$ for the discrete series is summed with integer spacing.
$\alpha,\alpha'$ for the continuous series and $j$ for the discrete series 
should take the values so that they are compatible with the conservation
of $J^3_0$.

\appsection{Rewriting $W(j_a;m_a)$}
We discuss various expressions of $W(j_a;m_a)$ in appendix B.
First, we rewrite $I$ in (\ref{I1}). 
Applying the three-term relation (\ref{3G}) to $C^{12}$ 
in (\ref{Cab}) (with `$a$'$=k'-\gamma -\gamma'$),
we obtain
\eqb
    s(\beta)s(\gamma-\beta') C^{12} &=& 
   s(\gamma) s(\gamma+\gamma'-k') C^{13} -s(\alpha)s(\beta')C^{21} 
   \period \nn
\eqe
A similar expression for $C^{31}$ is also obtained from (\ref{CbaCab}). 
These allow us to express $P^{12}$ in terms of $C^{12}$ and $C^{21}$:
\eqb
  (i/2)^2 P^{12} &=& D_1 C^{12} + D_3 C^{21} 
  \comma
\eqe
where
\eqb
  D_{1} &=& {s(\alpha')s(\beta) \o s(\gamma)s(\gamma')s(\gamma+\gamma'-k')}
    \lmb s(\alpha)s(\alpha')s(\gamma) 
   - s(\gamma')s(\alpha-k')s(\gamma -\beta') \rmb \comma  \nn \\
 D_3 &=& - {s(\alpha)s(\alpha')s(\beta)s(\beta')s(\gamma+\gamma') \o
     s(\gamma)s(\gamma')s(\gamma+\gamma'-k')}
  \period 
\eqe
The corresponding expression for $P^{21}$ is given by (\ref{CbaCab}). 
Hence, the integral $I$ becomes
\eqb
  I &=& D_{1} C^{12}[\alpha_i,\alpha'_i] C^{12}[\alphabar_i,\alphabar'_i]
     + D_{2} C^{21}[\alpha_i,\alpha'_i] C^{21}[\alphabar_i,\alphabar'_i]
   \label{I2} \\
  && \qquad \quad + \,
    D_3 \lmb C^{12}[\alpha_i,\alpha'_i] C^{21}[\alphabar_i,\alphabar'_i]
        + C^{21}[\alpha_i,\alpha'_i] C^{12}[\alphabar_i,\alphabar'_i]
    \rmb
  \comma \nn
\eqe
with $D_2 = D_1(\alpha_i \leftrightarrow \alpha'_i)$.
From (\ref{I2}), we find that $I$ is symmetric with respect to 
$(\alpha_i, \alpha'_i)$ and $ (\alphabar_i, \alphabar'_i)$, as it should be, 
when $\alpha_i-\alphabar_i, \alpha'_i-\alphabar'_i \in \bfZ$.
 
Using the above result, $W(j_a;m_a)$ is rewritten as in (\ref{Wjm2}).
Furthermore, $W(j_a;m_a)$ is expressed in various ways, since $C^{ab}$
can also take various forms  
because of, e.g., the identities (\ref{2G}). Which form is useful
depends on each discussion. Thus, we list several ones in addition to
those in (\ref{PCC}):
\eqb
   C^{12} 
   &=& {\Gamma(-N) \Gamma(-j_{13}) \Gamma(1+j_2+m_2) \Gamma(1+j_3-m_3) 
  \o \Gamma(1+j_{23}) \Gamma(-j_1-m_1) \Gamma(-j_3-m_3)}
\G{-j_1-m_1,-j_{12},1+j_{23}}{-2j_1,-j_1+j_3+m_2+1} \nn \\
   &=& 
  {\Gamma(-N) \Gamma(-j_{12}) \Gamma(-j_{13})\Gamma(1+j_2+m_2)  
  \o \Gamma(-j_1-m_1) \Gamma(-j_1+m_1)\Gamma(-j_3-m_3)} 
   \G{-j_3-m_3,1+j_3-m_3,-j_1+m_1}{-j_1 +j_2-m_3+1,-j_1-j_2-m_3}
   \comma \nn \\
   C^{31} 
      &=& {\Gamma(-j_{12}) \Gamma(1 + j_1 + m_1) \Gamma(1 + j_2 - m_2) 
               \Gamma(1 + j_3 + m_3)  \o
                \Gamma(1+N) \Gamma(1+j_{23}) \Gamma(-j_2-m_2) } 
       \label{Cs} \\ 
    && \quad \times \
  \G{1+N,1+j_{23},1+j_3-m_3}{j_2+j_3+m_1+2,2+2j_3}
   \period \nn
\eqe
For $C^{12}$, $u$ are 
$-j_3-m_3$ and $-j_1 -m_1$ in the first and second lines, respectively, 
whereas it is $-j_2-m_2$ for $C^{31}$. 
Similar expressions for $C^{21}$ and $C^{13}$
are obtained through (\ref{CbaCab}).   

\appsection{Analytic structure of $F$}
Here, we summarize the analytic structure of $ \F{a,b,c}{e,f} $.
It may be instructive to first consider $ \2F1 (a,b;c;1) $ defined by
\eqb
 \2F1 (a,b;c;1)  
   &=& {\Gamma(c) \o \Gamma(a) \Gamma(b)}\sum_{n=0}^\infty 
   {\Gamma(a+n) \Gamma(b+n) \o \Gamma(c+n) \Gamma(n+1)} 
  \period
\eqe
This has obvious poles at $ c \in \bfZ_{\leq 0}$.
This series is convergent only if  Re$ \, (c-a-b) > 0$, but
it can be analytically continued to the region Re$ \, (c-a-b) \leq  0$, e.g., 
by using its integral representation. One then expects that  
$ \2F1 (a,b;c;1) $ develops additional poles at $ c-a-b \in \bfZ_{\leq 0}$.
This is consistent with Gauss' formula: 
$\2F1 (a,b;c;1) = {\Gamma(c) \Gamma(c-a-b) \o \Gamma(c-a) \Gamma(c-b)} $.
Note that the residues can vanish in special cases because of the 
denominator.

Now, let us turn to the case of $ \F{a,b,c}{e,f} $, which is defined by
\eqb
  \F{a,b,c}{e,f} 
    &=& {\Gamma(e) \Gamma(f)\o \Gamma(a) \Gamma(b) \Gamma(c)}\sum_{n=0}^\infty 
   {\Gamma(a+n) \Gamma(b+n) \Gamma(c+n)\o \Gamma(e+n) \Gamma(f+n)\Gamma(n+1)}
   \period
\eqe
Note that this series generically reduces to a finite sum with 
$(n+1)$ terms when $a,b$ or $c$ takes a non-positive integer $-n$.
A consideration similar to the above shows that
the series is convergent only if Re$ \, u > 0$, 
and generically  
\eqb
 \F{a,b,c}{e,f} & \mbox{has simple poles at} & 
   e \comma f \comma u  \ \in \ \bfZ_{\leq 0} 
  \period
\eqe
This indeed agrees with the poles appearing in the integral
expression of $C^{ab}$ in \cite{Fukuda-H}. We remark that 
the poles coming from $ u \in \bfZ_{\leq 0} $ are absent when
the series reduces to a finite sum. 

From the above discussion, it follows 
that: (i) in (\ref{PCC}), $F$'s in $C^{12}$ and in a similar expression 
for $C^{21}$ are regular when $j_1+m_1 \in \bfZ_{\geq 0}$ and 
other parameters are generic; (ii) in (\ref{W1}),
$ { \Gamma(1+j_3-m_3) \o \Gamma(1+j_3-m_3 -n_1) }
   \F{-n_1,-j_{12},1+j_{23}}{-2j_1,1+j_3-m_3-n_1} $
is regular for generic $j_1$;  (iii) in (\ref{Cs}),
$ F $'s in the second expression of $C^{12}$, in 
$C^{31}$ and in similar expressions for $C^{21},C^{13}$ are regular
for generic $ j_1,j_2 \in \calP_+$ and $(j_3,m_3) \in  \calD^\pm_j$ 
or $\calC^\alpha_j$.

Although we do not have the formula for $ \3F2 $ corresponding to 
Gauss' formula, simple expressions of $ \3F2 $ in terms of 
the gamma functions are known in special cases. In particular, 
from Saalschutz's theorem, we obtain \cite{Slater} 
\eqb
  \F{a,b,c}{e,f} &=& { \Gamma(e) \Gamma(1+a-f) \Gamma(1+b-f) \Gamma(1+c-f)
                     \o \Gamma(1-f) \Gamma(e-a) \Gamma(e-b) \Gamma(e-c) }
   \comma \label{Saalschutz} 
\eqe 
when $u = 1$ and one of $a,b,c$ is a negative
integer $-n$. This reduces to Gauss' formula for $n \to \infty$. 
Applying (\ref{Saalschutz}) to $F$ in $C^{12}$ gives 
\eqb
    && { \Gamma(1+j_3-m_3) \o \Gamma(1+j_3-m_3 -n_1) }
   \F{-n_1,-j_{12},1+j_{23}}{-2j_1,1+j_3-m_3-n_1}  
  =
    {\Gamma(1+2j_1-n_1) \Gamma(1+j_{13}) \Gamma(1+N)
        \o \Gamma(1+2j_1) \Gamma(m_2-j_2) \Gamma(1+j_2+m_2)}
   \comma \nn \\
  &&  \label{SaalschutzF} 
\eqe
for $-j_3-m_3 = 1$, $j_1+m_1=n_1$ and $\Sigma m_a = 0$.
%

%
%\newpage
\appsection{From $W_1^0$ to $W_1$}
In appendix D, we sketch the derivation of (\ref{W1}) by acting with $J_0^a$, 
$\Jbar_0^a$ on (\ref{W10}). We follow the argument in \cite{Becker-B}. 
First, we note an identity 
\eqb
  \Bbra \, e^{\alpha J^+_0} \Phi^{j_1}_{-j_1 \, -j_1} e^{-\alpha J^+_0} 
  \Phi^{j_2}_{m_2\mbar_2}  \Phi^{j_3}_{m_3\mbar_3} \Bket
 &=& 
  \Bbra  \Phi^{j_1}_{-j_1 \, -j_1} e^{-\alpha J^+_0} 
  \Phi^{j_2}_{m_2\mbar_2}  \Phi^{j_3}_{m_3\mbar_3} \, e^{\alpha J^+_0} \Bket
  \period
\eqe
This gives
\eqb
  && \sum_{n_1 = 0}^\infty \frac{(-\alpha)^{n_1}}{n_1 !} 
  {\Gamma(2j_1+1) \o \Gamma(2j_1+1-n_1)} 
   \Bbra \Phi^{j_1}_{-j_1+n_1 \, -j_1} \Phi^{j_2}_{m_2\mbar_2} 
   \Phi^{j_3}_{m_3\mbar_3} \Bket
     \label{ladder} \\
  && = \ \sum_{n_2,n_3 = 0}^\infty \frac{\alpha^{n_2+n_3}}{n_2 ! n_3 !} 
  {\Gamma(j_2-m_2+1) \o \Gamma(j_2-m_2+1-n_2)} 
  {\Gamma(j_3-m_3+1) \o \Gamma(j_3-m_3+1-n_3)} 
   \Bbra \Phi^{j_1}_{-j_1 \, -j_1} \Phi^{j_2}_{m_2+n_2 \, \mbar_2} 
   \Phi^{j_3}_{m_3+n_3 \, \mbar_3} \Bket
  \period \nn
\eqe
Equating the terms on both sides with the same power in $\alpha$, 
we obtain
the expression of the three-point functions with $ n_1 \neq 0 $ in terms of
that with $n_1 = 0$. Taking into account the right sector, 
we find that
\eqb
   \Bbra \Phi^{j_1}_{-j_1+n_1 \, -j_1+\nbar_1} \Phi^{j_2}_{m_2\mbar_2} 
   \Phi^{j_3}_{m_3\mbar_3} \Bket
  &=& (2\pi)^2 \delta^2(-j_1+n_1+m_2+m_3) 
    \, \calC_1(j_a,m_a) {\calC}_2(j_a,\mbar_a) \nn  \\
   && \quad \ \times 
    \pi^2 {\Delta(-N) \Delta(2j_1+1) \o \Delta(1+j_{12}) \Delta(1+j_{13})}
    D(-j_a-1) \comma  \label{calCCD}
\eqe
where
\eqb
  \calC_1(j_a,m_a) &=& \frac{\Gamma(-2j_1)}{\Gamma(m_1-j_1)}
    \frac{\Gamma(j_2+m_2+1)}{\Gamma(j_3-m_3+1-n_1)}
    \frac{\Gamma(j_3-m_3+1)}{\Gamma(-j_3-m_3-n_1)} \nn \\
    && \quad \times \  
   \F{-j_1-m_1,-j_2+m_2,j_2+m_2+1}{-j_1+j_3+m_2+1,-j_1-j_3+m_2}  
     \label{calC} \\
   &=& 
   \frac{\Gamma(j_2+m_2+1)}{\Gamma(-j_3-m_3)}
   \frac{\Gamma(j_3-m_3+1)}{\Gamma(j_3-m_3+1-n_1)}
   \F{-j_1-m_1,1+j_{23},-j_{12}}{-2j_1,1+j_3-m_3-n_1} \comma \nn  \\
  {\calC}_2(j_a,\mbar_a) &=&  {s(j_2+\mbar_2) \o s(j_3+\mbar_3+\nbar_1)}
   \calC_1(j_a,\mbar_a)
  \period \nn
\eqe
In obtaining $\calC_1$, we have used 
$ {\Gamma(\alpha) \o \Gamma(\alpha-n)} 
  = (-)^n {\Gamma(1-\alpha+n) \o \Gamma(1-\alpha)} $ repeatedly
 for the first line, and the second formula 
 in (\ref{2G}) for the second line. Comparing (\ref{calCCD}) with
(\ref{Phijm3pt}), we confirm that $W_1$ is correctly recovered.  

In \cite{Becker-B}, the quantity corresponding to
$\calC_1\calC_2$ in (\ref{calCCD}) is $[{\calC}_2(j_a,m_a)]^2$.  
This is because 
the quantity corresponding to $W_1^0$ 
has a different phase (in addition to $m_a = \mbar_a$).

%
%\newpage
%%%%%%%%%%%%%%%%%%%%%%%%%%%%%
% references
%%%%%%%%%%%%%%%%%%%%%%%%%%%%%
\def\thebibliography#1{\list
 {[\arabic{enumi}]}{\settowidth\labelwidth{[#1]}\leftmargin\labelwidth
  \advance\leftmargin\labelsep
  \usecounter{enumi}}
  \def\newblock{\hskip .11em plus .33em minus .07em}
  \sloppy\clubpenalty4000\widowpenalty4000
  \sfcode`\.=1000\relax}
 \let\endthebibliography=\endlist
%%%%%%%%%%%%%%%%%%%%%%%%%%%%%
%\vskip 8ex
\newpage
\begin{center}
 {\bf References}
\end{center}
\vskip 0ex
%\par\smallskip

%
\end{document}